# Use of Wedge Absorbers in MICE

David Neuffer[1], D. Summers[2], T. Mohayai[1,3], P. Snopok [1,3], C. Rogers[4]

[1]Fermilab, PO Box 500, Batavia Il 60510
[2]Unversity of Mississippi, Oxford, MS 38655
[3]IIT, Chicago 60166
[4]RAL, Didcot, England OX11OQX

**Abstract.** Wedge absorbers are needed to obtain longitudinal cooling in ionization cooling. They also can be used to obtain emittance exchanges between longitudinal and transverse phase space. There can be large exchanges in emittance, even with single wedges. In the present note we explore the use of wedge absorbers in the MICE experiment to obtain transverse–longitudinal emittance exchanges within present and future operational conditions. The same wedge can be used to explore "direct" and "reverse" emittance exchange dynamics, where direct indicates a configuration that reduces momentum spread and reverse is a configuration that increases momentum spread. Analytical estimated and ICOOL and G4BeamLine simulations of the exchanges at MICE parameters are presented. Large exchanges can be obtained in both reverse and direct configurations.

## INTRODUCTION

In muon collider scenarios, beam cooling to emittances as small as 25μ (transverse, rms, normalized) are required to ensure high luminosity at multiTeV energies[1, 2]. Reduction of muon beam emittance is also necessary in producing beams suitable for neutrino factories, and can improve the intensity in other muon-based experiments.[3]

Ionization cooling can be used to reduce muon beam emittances, following the cooling equation for transverse, normalized emittances:

$$\frac{d\varepsilon_N}{ds} = -\frac{g_t}{\beta^2 E}\frac{dE}{ds}\varepsilon_N + \frac{\beta_\perp E_s^2}{2\beta^3 m_\mu c^2 L_R E} \quad (1)$$

where the first term is the frictional cooling effect and the second is the multiple scattering heating term. Here $L_R$ is the material radiation length, $\beta_\perp$ is the betatron focusing function, and $E_s$ is the characteristic scattering energy (~14 MeV), and $g_t$ is the transverse partition number. In the absence of transverse-longitudinal coupling (such as wedge absorbers) $g_t = 1$. The equilibrium emittance is:

$\varepsilon_{N,eq} = \frac{\beta_\perp E_s^2}{2\beta g_t m_\mu c^2 L_R \frac{dE}{ds}}$. For MICE, liquid $H_2$ and lithium hydride (LiH) are being used. Be and diamond (C) are also used in cooling scenarios, and for the MICE experiment we are considering using polyethylene (~$C_2H_4$) as a first wedge experiment material. For LiH the density is 0.82gm/cm$^3$, $L_R$ = 96.6cm, dE/dx = 1.73, and for poly the density is 0.94gm/cm$^3$, $L_R$ = 44.6cm, dE/dx = 2.16.[4] The equilibrium emittances are $\beta_t /(\beta g_t)$ times 0.0064 and 0.0091 for LiH and poly, respectively. Thus heating from multiple scattering is ~50% larger for poly, although energy loss is similar.

## Emittance Exchange with Wedge Absorbers

In final cooling, the transverse emittance is reduced while the longitudinal emittance is allowed to increase [5]. In recent scenarios the longitudinal heating is nearly equal to the transverse cooling, so the channel is effectively an emittance exchange system. The simplest form of emittance exchange is found by passing the beam through a wedge absorber, where the bunch width is transformed into an energy width. It was previously noted that large emittance exchanges by single wedges are possible near final cooling parameters.[6]

Figure 5 shows a stylized view of the passage of a beam with dispersion $\eta_0$ through an absorber. The wedge is approximated as an object that changes particle momentum offset $\delta = \Delta p/P_0$ as a function of $x$, and the wedge is shaped such that that change is linear in $x$. (The change in average momentum $P_0$ is ignored, in this approximation. Energy straggling and multiple scattering are also ignored.) The rms beam properties entering the wedge are given by the transverse emittance $\varepsilon_0$, betatron amplitude $\beta_0$, dispersion $\eta_0$ and relative momentum width $\delta_0$. (To simplify discussion the beam is focussed to a betatron and dispersion waist at the wedge: $\beta_0'$, $\eta_0' = 0$. This avoids the complication of changes in $\beta'$, $\eta'$ in the wedge.) The wedge is represented by its relative effect on the momentum offsets $\delta$ of particles within the bunch at position $x$:

$$\frac{\Delta p}{p} = \delta \to \delta - \frac{(dp/ds)\tan\theta}{P_0}x = \delta - \delta' x$$

$dp/ds$ is the momentum loss rate in the material ($dp/ds = \beta^{-1}dE/ds$). $x\tan\theta$ is the wedge thickness at transverse position $x$ (relative to the central orbit at $x=0$), and $\delta' = dp/ds \tan\theta/P_0$ to indicate the change of $\delta$ with $x$. The initial dispersion and the wedge can be approximated by linear transformations in $x$-$\delta$ phase space and the transformations are phase-space preserving. The dispersion is represented by the matrix: $\mathbf{M}_\eta = \begin{bmatrix} 1 & \eta_0 \\ 0 & 1 \end{bmatrix}$, since $x \Rightarrow x + \eta_0\delta$. The wedge can be represented by the matrix: $\mathbf{M}_\delta = \begin{bmatrix} 1 & 0 \\ -\delta' & 1 \end{bmatrix}$. The dispersion + wedge becomes: $\mathbf{M}_{\eta\delta} = \begin{bmatrix} 1 & \eta_0 \\ -\delta' & 1-\delta'\eta_0 \end{bmatrix}$. This matrix changes the dispersion, the momentum width, and the transverse beam size (dispersion removed). The momentum width is changed to:

$$\delta_1 = \sqrt{g_1\sigma_0\delta_0} = \delta_0\left[(1-\eta_0\delta')^2 + \frac{\delta'^2\sigma_0^2}{\delta_0^2}\right]^{1/2}.$$

The bunch length is unchanged. The longitudinal emittance is therefore changed by the factor $\delta_1/\delta_0$. The transverse emittance is changed by the inverse of this factor. The new values of $(\eta, \beta)$ are:

$$\eta_1 = -\frac{a_1}{g_1} = \frac{\eta_0(1-\eta_0\delta') - \delta'\frac{\sigma_0^2}{\delta_0^2}}{(1-\eta_0\delta')^2 + \delta'^2\frac{\sigma_0^2}{\delta_0^2}}, \text{ and } \beta_1 = \beta_0\left[(1-\eta_0\delta')^2 + \frac{\delta'^2\sigma_0^2}{\delta_0^2}\right]^{-1/2}.$$

As currently presented, the wedge exchanges emittance between one transverse dimension and longitudinal; the other transverse plane is unaffected. Serial wedges could be used to balance $x$ and $y$ exchanges, or a more complicated coupled geometry could be developed.

Wedge parameters can be arranged to obtain large exchange factors in a single wedge. In upstream systems the wedges can be arranged to obtain a factor of longitudinal cooling (at expense of transverse heating). In final cooling for a lepton collider, we wish to reduce transverse emittance at the cost of increased longitudinal emittance.

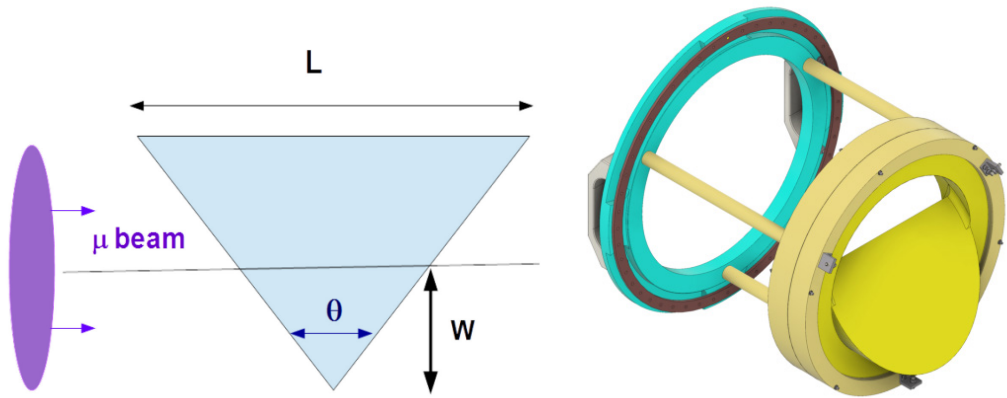

**Figure 1.** Schematic view of a muon beam passing through a wedge (left). Wedge as will be constructed for MICE (right), designed to fit in the same absorber

## Thick Wedges at Final Cooling Parameters

For final cooling, the beam and wedges should be matched to obtain a large factor of increase in momentum spread.[7, 8] That means that the energy spread induced by the wedge should be much greater than the initial momentum spread: $\delta_0 << \delta'\sigma_0 = \frac{2\tan\left(\frac{\theta}{2}\right)\frac{dp}{ds}}{P_0}\sigma_0$. Thus the incident beam should have a small momentum spread and small momentum $P_0$ and the wedge should have a large $\tan(\theta/2)$, large $dp/ds$ and a large $\sigma_0 = (\varepsilon_0\beta_0)^{1/2}$. ($\varepsilon_0$ is unnormalized, rms in this section.) Beam from a final cooling segment (high-field solenoid or Li lens) is likely to have $P_0 \approx$ 100—150 MeV/c, and $\delta p \approx$ 3MeV/c. For optimum single wedge usage, $\delta p$ should be reduced to ~0.5MeV/c, and this can be done by rf debunching of the beam to a longer bunch length.

For maximal wedge effect, the beam size should be matched to the wedge size $w$ ($w \approx 2\sigma_0$). To minimize multiple scattering heating at the wedge, $\beta_0$ should be small (< a few cm) and the wedge should be a low-Z material, but should also be high density.

At final cooling parameters ($\beta_0 \approx$ 1cm, $\varepsilon_0 \approx$ 100μ), $\sigma_0 \approx$ 1mm, which means the final cooling wedges would be only a few mm in size. As an example we consider a high-density C wedge (diamond density) with an input μ beam at 100 MeV/c, $\delta E$ =0.5 MeV ($\delta p$ =0.73 MeV/c), $\varepsilon_x = \varepsilon_y$ =0.013cm, matched to $\beta_t$ = 2cm at the center of the wedge ($\sigma_x$=1.65mm), and zero initial dispersion. The wedge parameters are: $w$=3mm, $\theta$=85° (~5.6mm thick at beam center).

ICOOL simulation results are presented in Table 1 and figures 2 and 3, with emittances calculated using EMITCALC (part of the ICOOL tools). The wedge-plane (x) emittance is reduced to 0.0025cm (a factor of 5.2 !) while the y-plane emittance is unaffected. However, the energy spread is increased to ~3.9MeV, diluting 6-D emittance by ~50%. Some of this dilution may be correctable by improving the phase-space match and using higher order correction. In any case, this simple wedge obtains a transverse emittance below the goals of a high-energy collider.

**Table 1**: Beam parameters at entrance, center and exit of a w=3mm, θ=85° diamond wedge. The z = 0, 0.6, 1.2cm rows are beam parameters before, at the center, and after the wedge. The 0.6cm values also indicate results that can be obtained with a half-strength wedge.

| z(cm) | $P_z$(MeV/c) | $\varepsilon_x(\mu)$ | $\varepsilon_y(\mu)$ | $\varepsilon_L$ (mm) | $\sigma_E$(MeV) | 6-D ε increase |
|---|---|---|---|---|---|---|
| 0 | 100 | **129** | 127 | 1.0 | 0.50 | 1.0 |
| 0.6 | 95.2 | **40.4** | 130 | 4.03 | 1.95 | 1.29 |
| 1.2 | 90.0 | **25.0** | 127 | 7.9 | 3.87 | 1.54 |

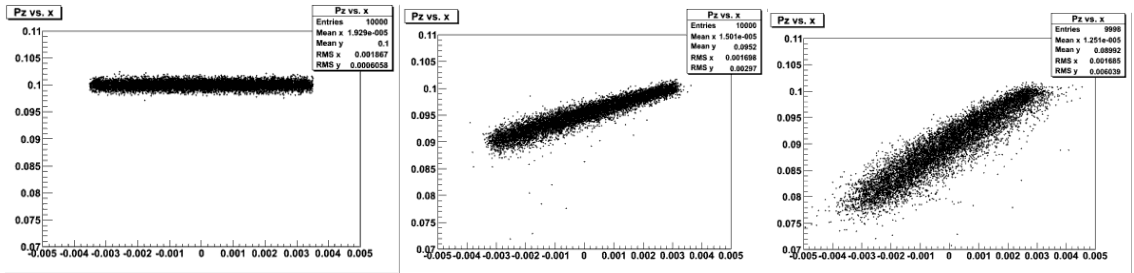

**Figure 2**. $x$-$P_z$ distributions for μ beam passing through a wedge system, shown at entrance, middle and end of the wedge from ICOOL simulation).

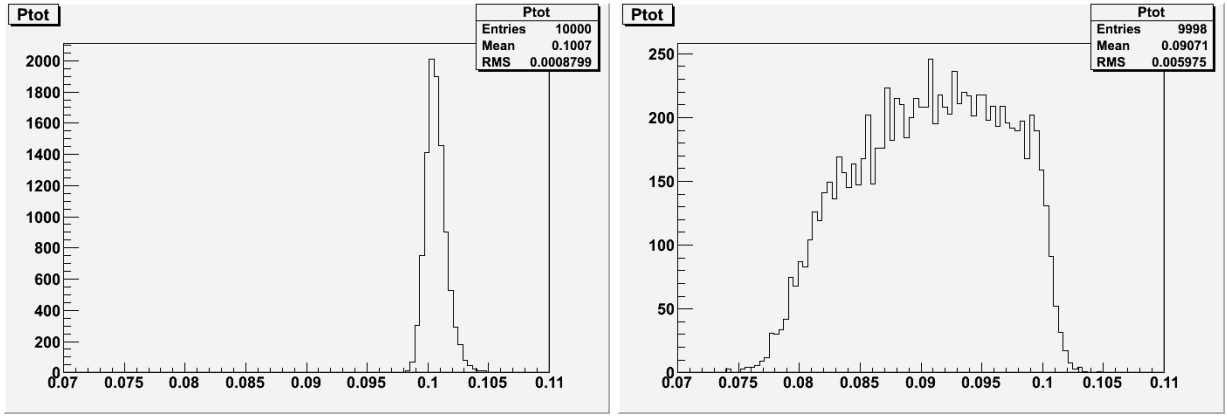

**Figure 3.** Momentum spread before and after a diamond wedge for final cooling.

The beam from a wedge transform could be reaccelerated and phase-energy rotated to a longer bunch with small δE for a pass through a second wedge. If properly rematched, ICOOL simiulations indicate that the second wedge could reduce vertical emittance to ~0.0025mm, while horizontal emittance increase could be limited to keep $\varepsilon_x$ < 0.03mm (30μ).

## Wedge Experiment in MICE

The MICE experiment[11] has considered inserting a wedge absorber into the beam line for measurements of emittance exchange cooling.[12] The MICE experiment has considerable flexibility in beam definition, and can obtain an initial beam with small δp by selecting particle tracks from the ensemble measured in MICE within that momentum band. Passage of that beam through a wedge will cause "reverse" emittance exchange in which longitudinal emittance increases while transverse emittance decreases, and large exchange factors can be obtained. [13]

In the same experiment an initial beam with a dispersion at the wedge can be selected from the ensemble of measured particle tracks, with the dispersion oriented such that the wedge reduces the beam momentum spread. This "direct" emittance exchange would be a configuration that could cool the beam longitudinally, and would be a demonstration of the longitudinal cooling effect that is needed to obtain simultaneous transverse and longitudinal cooling in a complete cooling channel.

The MICE step 4 cooling channel consists of an upstream spectrometer solenoid module, a focus coil/absorber module and a downstream spectrometer solenoid module. The solenoid modules include a coil triplet that would contain the spectrometer within an approximately constant magnetic field, and two matching coils for matching the beam optics into the focus/absorber module. The focus/absorber contains two coils with fields that are adapted to obtain a strong beam focus at the center of the focus coil module, between the two coils. The cooling absorbers are placed between the coils. The focus coils can have opposite polarities (flip mode) or the same polarity (non-flip) mode.

Within the present step 4 configuration (see fig. 4 and 5) absorbers are placed between the focus coils. Currently (2/2017) a LiH slab with ~6.5cm thickness is placed in the center, and a later run would insert a liquid $H_2$ absorber. For a wedge absorber experiment, we would use a polyethylene (~$C_2H_4$) wedge, placed in the same location and designed to use the same absorber holder as the LiH slab. [fig. 1, right] A polyethylene absorber can be obtained and machined at low cost in a short amount of time, without material-related safety problems. Performance would be similar to a LiH wedge, but with a bit more multiple scattering.

Beam optics in the MICE channel has been compromised by the failure of one of the downstream match coils, and it is currently operating with both downstream match coils off. Current settings matched to these optics have been developed. We first use one of these configurations, labelled ( 2016-05 3), which was designed for 200MeV/c beam. Fig. 5 shows the magnetic field on axis. Fig. 6 shows the betatron function for a 200 MeV/c, matched to $\beta_t$ = 0.44m within the 3T upstream solenoid. At the absorber, $\beta_t \cong 0.77$m. Large betatron fluctuations following the absorber into the downstream solenoid occur because of the missing match coils.

We use a polyethylene absorber with $w$=6.5cm, $\theta$=45°, with the wedge oriented along $y$. (A Be or LiH wedge would have superior performance, but greater expense, and would not greatly improve the initial proof of principle demonstration.)

A maximal exchange is obtained by matching the beam size to the wedge size. For ~gaussian beams this means a beam size of $2\sigma \cong w$. For a 200 MeV/c beam and an emittance of $\varepsilon_N$ = 0.003 m, $\sigma$ =~3.3cm, which is a reasonably close match to the wedge.

This configuration was simulated using G4BeamLine[14]. With an initial beam with no dispersion and $\delta p$ = 2 MeV/c, one finds in simulation that the energy spread increases from 1.83 to 6.68 MeV, a factor of ~3.6. The transverse eigen emittances change from 2.93 and 2.77mm to 3.05 and 1.02 mm, indicating a reduction of one of the emittances by a factor of 3. The eigen emittances were calculated using EMITCALC and a similar evaluation could be obtained on the measured beams in the MICE spectrometers. Note that the eigenmode emittances remain separated throughout the transport from absorber to spectrometer, and should be measureable within the spectrometer. The increase in emittance for both modes seen for z > 2.0 m is a result of the optical mismatch caused by the match coil failure, and is also seen in the non-wedge cooling measurements.

Simulation results are displayed graphically in figures 8 and 9. In figure 8 the momentum distributions before and after the foil are displayed. The change in momentum spread is similar in magnitude to the final focusing emittance exchange discussed above. The momentum distributions are readily measureable in the MICE distribution and this measurement in MICE would be a graphic demonstration of a scale model of the final cooling emittance exchange discussed above.

With the same initial data sample, but changing the initial $\delta p$ selection, the same experiment could map out the emittance exchange effect as a function of momentum, and identify the optimum $\delta p$ for maximal exchange.

Note that both the momentum increase and the transverse emittance decrease are very large effects and should be relatively easy to measure in MICE. These measurements will complement the more difficult measurements of emittance reduction, which have been made more difficult by the matching coil failure.

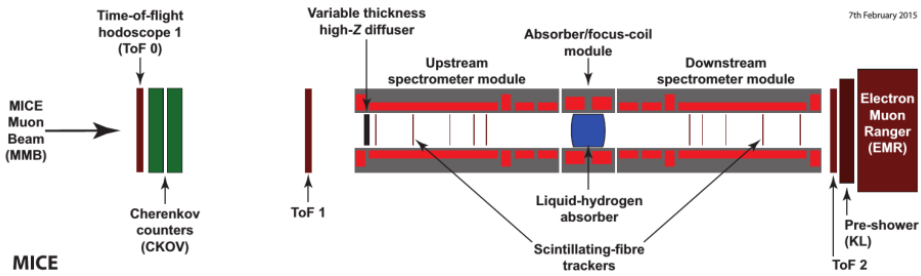

**Figure 4:** Schematic overview of the MICE step 4 configuration showing beam detector components before and after the cooling channel assembly that includes the spectrometer solenoids.

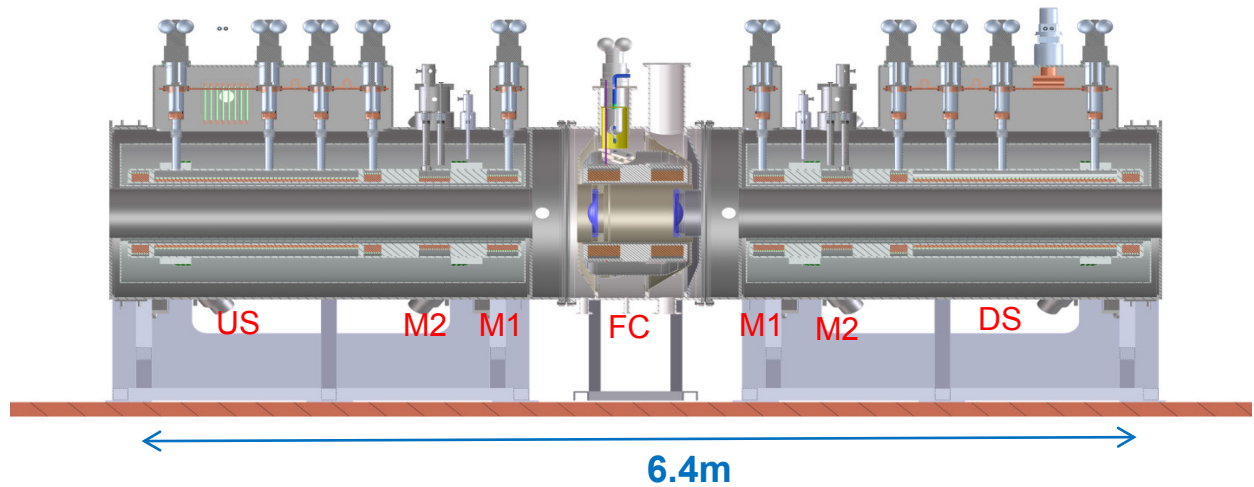

**Figure 5:** CAD drawing of the cooling channel, showing the cryomodules, and the magnetic coils (brown). The match coils are labelled M2 and M1. The downstream M1 coil is non-functional.

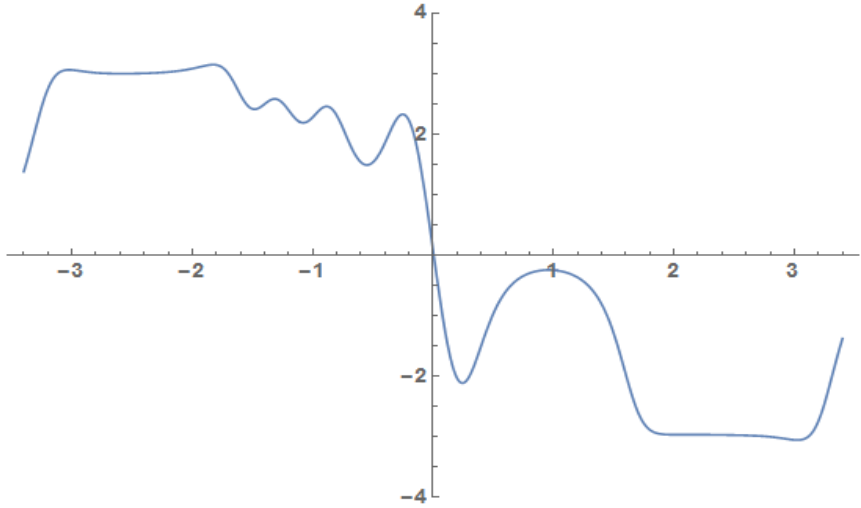

**Figure 6:** Magnetic field ($B_z$) in the MICE step 4 configuration with 2016-05v3 currents in the coils.

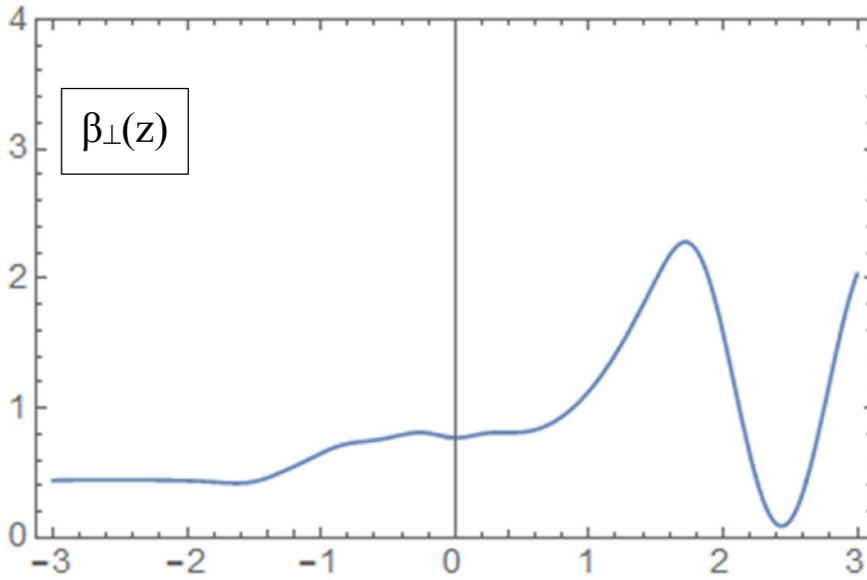

**Figure 7:** Betatron function $\beta_t(z)$ in m for 200 MeV/c beam in the MICE cooling channel, matched to $\beta_t=0.44$m in the upstream solenoid (~-3 to -2 m).

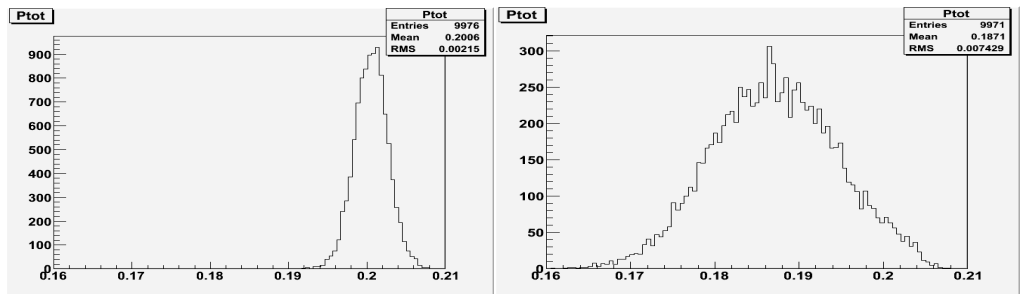

**Figure 8:** Momentum distribution before and after absorber

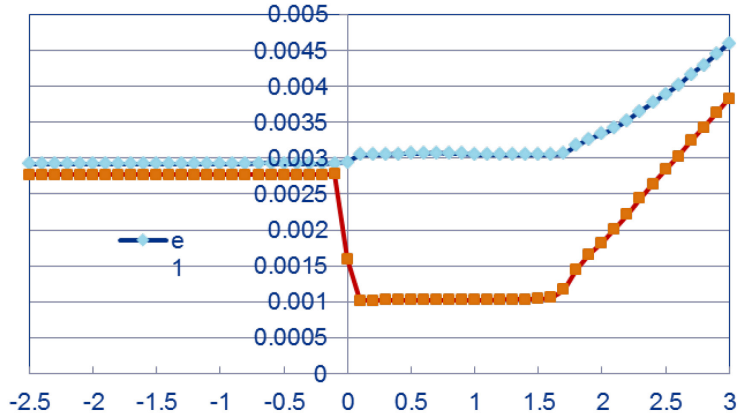

**Figure 9:** Development of the transverse eigenemittances from the upstream solenoid into the downstream solenoid, as calculated in G4BeamLine simulations. Initial beam was at 200 MeV/c with a width of ~2%. Emittance exchange reduces one emittance by a factor of 3 while the other is slightly heated (consistent with the multiple scattering effect). The emittances are preserved throughout the transport into downstream solenoid, where the mismatched dynamics causes beam filamentation.

### Direct emittance exchange example

The ensemble of particle tracks entering the MICE cooling channel includes a momentum spread of ~10%. Within this larger momentum spread a dispersion aligned with the wedge can be generated at the absorber and that dispersive beam could have its momentum spread reduced by the wedge, and this would imply longitudinal cooling. 6-D phase space ionization cooling requires such cooling. Within the single wedge passage in MICE, the effect is predominantly emittance exchange, so transverse emittance must increase, in the plane of the wedge and dispersion.

This process can also be observed at MICE. For an example we can use the same MICE magnet configuration used above, and the same transverse emittance. However, with dispersion the beam size would become larger, and would not be matched to the 45º wedge. 6.5cm wedge. For a numerical example we consider a 30º, w=10cm wedge. We introduce a matched $\varepsilon_t$=0.003m beam, but with an intial dispersion and $\sigma_E$ = 8MeV, and simulate using G4Beamline.

An initial dispersion of ~0.75m is ~8 MeV at the absorber. The absorber reduces the energy spread to 5.9 MeV, which would indicate longitudinal cooling. The transverse eigenmodes change to 0.0028 and 0.0053 m emittances; the transverse heating in the vertical mode seems to be a bit larger than the longitudinal cooling.

This change in momentum spread should be readily measureable at MICE, although it is not as large an effect as the reverse emittance exchange case. Transverse emittance changes should also be measureable. The general purpose of measuring longitudinal emittance reduction with a wedge absorber should be achievable.

The optics is not as easily matched, since it requires an initial dispersion and this first simulation attempt, which uses ICOOL to generate the initial beam and G4beamline for the tracking, is not optimally matched. Future studies will provide better matched cases, which should reduce the transverse emittance dilution.

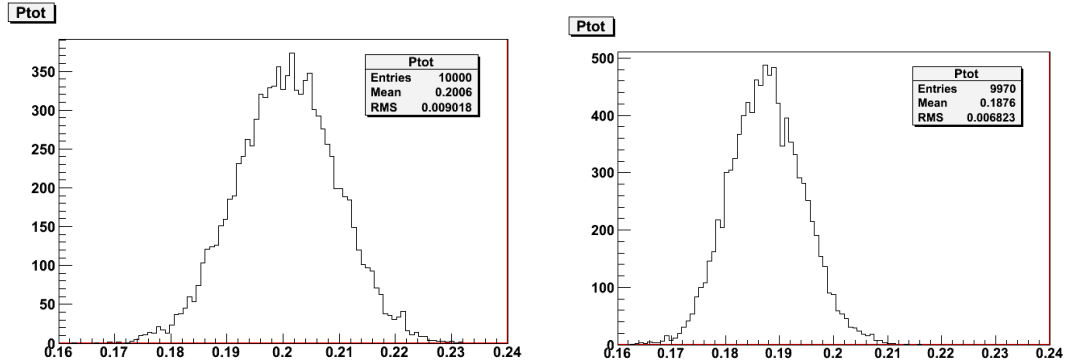

**Figure 10:** Energy spread before and after direct emittance exchange. The average momentum is reduced from 200.6 to 187.6 MeV/c while the momentum spread decreases from 9.02 to 6.82 MeV/c

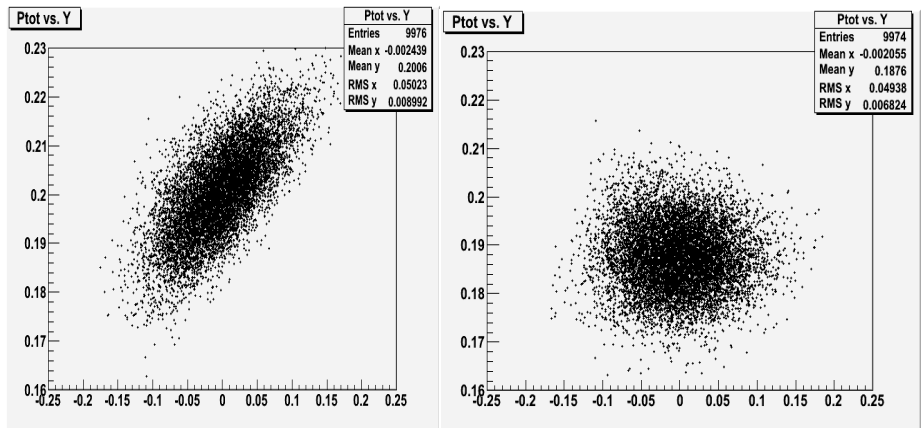

**Figure 11:** Y-P phase space before and after the wedge. The vertical dispersion (Y-P correlation) is greatly reduced, while the momentum spread decreases. The rms beam size is ~5cm and is matched to a w=10 cm wedge.

## Summary

We have discussed the importance of wedge absorbers in ionization cooling, emittance exchange, and beam phase manipulations. We note that the MICE experiment presents an important opportunity for a beam based measurement of these effects, and the measured effects both in direct and reverse emittance exchange can be large. A simple polyethylene (plastic) wedge can be designed and inserted in the MICE transport channel to measure these effects.